\documentclass[journal, twocolumn]{IEEEtran}
\usepackage{amsmath,amssymb,graphicx,epsfig,cite,fancyhdr,amsthm}
\usepackage{hyperref}
\usepackage{tikz}
\usetikzlibrary{calc}
\usepackage{pdfsync}
\usepackage{bm}
\usepackage{epstopdf}
\newcommand{\bfsub}[2]{\mbox{\boldmath$#1$}$_{#2}$}
\newcommand{\bfsubp}[3]{\mbox{\boldmath$#1$}$_{#2}^{#3}$}
\newcommand{\mybf}[1]{\mbox{\boldmath$#1$}}
\makeatletter
\newcommand*\dashline{\rotatebox[origin=c]{90}{$ \dabar@\dabar@\dabar@\dabar@\dabar@\dabar@\dabar@$}}
\makeatother

\theoremstyle{remark}

\theoremstyle{definition}

\theoremstyle{example}

\begin{document}

\title{Distributed Lossy Source Coding Using Real-Number Codes}


\author{\IEEEauthorblockN{Mojtaba Vaezi and Fabrice Labeau}\\
\IEEEauthorblockA{Department of Electrical and Computer Engineering\\
McGill University, Montreal, Quebec H3A 2A7, Canada\\
Email: mojtaba.vaezi@mail.mcgill.ca, fabrice.labeau@mcgill.ca}
}


\maketitle

\begin{abstract}
We show how real-number codes can be used to compress correlated
sources, and establish a new framework for distributed lossy source
coding, in which we quantize compressed sources
instead of compressing quantized sources. This
change in the order of binning and quantization blocks makes it
possible to model correlation between continuous-valued sources
more realistically and correct quantization error when the sources
are completely correlated. The encoding and decoding procedures
are described in detail, for discrete Fourier transform (DFT)
codes. Reconstructed signal, in the mean-squared error sense,
is seen to be better than or close to quantization error level in
the conventional approach.
\end{abstract}


\IEEEpeerreviewmaketitle

\section{Introduction}\label{sec:intro}

\let\thefootnote\relax\footnotetext{This work was supported by Hydro-Qu\'{e}bec, the Natural Sciences and Engineering Research Council of Canada and McGill University in the framework of the NSERC/Hydro-Qu\'{e}bec/McGill Industrial Research Chair in Interactive Information Infrastructure for the Power Grid.

%
%
}

The distributed source coding (DSC) deals with compression of correlated
sources which do not communicate with each other \cite{SW}.
Lossless DSC
(Slepian-Wolf coding), has been realized by different binary channel codes,
including LDPC \cite{liveris2002compression} and turbo codes \cite{aaron2002compression}.
The Wyner-Ziv coding problem \cite{WZ}, deals with lossy
data compression with side information at the decoder,
under a fidelity criterion.
Current approach in the DSC of a continuous-valued source is to first convert it
 to a discrete-valued source using quantization, and then to apply Slepian-Wolf coding
 in the binary field. Similarly, a practical Wyner-Ziv encoder is realized by
 cascading a quantizer and Slepian-Wolf encoder \cite{Girod,Xiong}.
 In other words, the quantized source is compressed. There are, hence, source coding
 (or quantization) loss and channel coding (or binning) loss.
 This approach is based on the assumption that there is still
 correlation remaining in the quantized version of correlated sources.

In this paper, we establish a new framework for the Wyner-Ziv coding.
We propose to first compress the continuous-valued source and then
quantize it, as opposed to the conventional approach. The
compression is thus in the real field, aiming at
representing the source with fewer samples.

To do compression, we
generate either syndrome or parity samples of the input sequence using a real-number
channel code, similar to what is done to compress a binary sequence
of data using binary channel codes. Then, we quantize these syndrome or parity
samples and transmit them. There are still coding (binning)
and quantization losses; however, since coding is performed before
quantization, error correction is in the real field and quantization
error can be corrected when two sources are completely correlated
over a block of code. A second and more important advantage of this
approach is the fact that the correlation channel model can be more
realistic, as it captures the correlation between continuous-valued
sources rather than quantized sources. In the conventional approach,
it is implicitly assumed that quantization of correlated signals
results in correlated sequences in the discrete domain which is
not necessarily correct due to nonlinearity of quantization operation. In addition,
most of previous works assume that this correlation,
in the binary field, can be modeled by a binary symmetric channel
(BSC) with a known crossover probability.
To avoid the loss due to inaccuracy of correlation model, we
exploit correlation between continuous-valued sources before quantization.

Specifically, we use real BCH-DFT codes \cite{Marshall}, for compression in the real field. Owing to
the DFT codes, the loss due to quantization can be decreased by a
factor of $k/n$ for an $(n,k)$ DFT code \cite{goyal2001quantized}, \cite{rath2004frame}. Additionally, if the two
sources are perfectly correlated over one codevector,
reconstruction loss vanishes. This is achieved in view of modeling the correlation between
the two sources in the continuous domain. Finally, the proposed scheme seems more suitable
for low-delay communication because using short DFT codes a reconstruction error better
than quantization error is achievable.

The rest of this paper is organized as follows. In Section~\ref{sec:sys},
we motivate and introduce a new framework for lossy DSC. In
Section~\ref{sec:DFT}, we briefly review encoding and decoding in real
DFT codes. Then in Section~\ref{sec:WZ}, we present the DFT encoder
and decoder for the proposed system, both in the syndrome and parity approaches.
These two approaches are also compared in this section. Section~\ref{sec:sum}
discusses the simulation results. Section~\ref{sec:con} provides our concluding remarks.

\section{Proposed System and Motivations}\label{sec:sys}
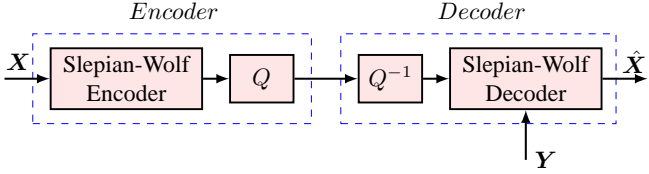
\begin{figure}[!htb]
\begin{center}
\scalebox {.9}{
\begin{tikzpicture}
[auto, block/.style ={rectangle, draw=blue, thick, fill=blue!10, text width=5em,align=center, rounded corners, minimum height=2em},
 block1/.style ={rectangle, draw=black, thick, fill=red!10, text width=6.5em,align=center, minimum height=2em},
 block2/.style ={rectangle, draw=white, thick, fill=white, text width=2em,align=center, rounded corners, minimum height=1em},
 line/.style ={draw, thick, -latex',shorten >=2pt},
 line1/.style ={draw, thick, -.latex',shorten >=2pt},
 cloud/.style ={draw=red, thick, ellipse,fill=red!20,
 minimum height=1em}]
 \draw(3.7,-1)node[block2,text width=6em] {$Encoder$};
 \draw(8.25,-1)node[block2,text width=6em] {$Decoder$};
 \draw(1.4,-1.75)node[block2,text width=1em] {$\bm{X}$};
 \draw(10.5,-1.75)node[block2,text width=1em] {$\hat{\bm{X}}$};
\draw (3,-2) node[block1, text width=5.7em] (C) {Slepian-Wolf Encoder };
\draw (5.0,-2) node[block1, text width=2em] (C1) {$Q$};
\draw (6.9,-2) node[block1 , text width=2em] (C2) {$Q^{-1}$};
\draw (8.9,-2) node[block1, text width=5.7em] (C3) {Slepian-Wolf Decoder };
\draw[blue,dashed]  ($(C.north west)+(-0.25,0.2)$) rectangle ($(C.south east)+(1.6,-0.2)$) ;
\draw[blue,dashed]     ($(C.north west)+(4.3,0.2)$) rectangle ($(C.south east)+(6.1,-0.2)$);
\draw [-latex] [line width=.3mm]  (1.2,-2) |- (C.west) ;
\draw [-latex] [line width=.3mm] (C.east) |- (C1.west) ;
\draw [-latex] [line width=.3mm](C1.east) |- (C2.west);
\draw [-latex] [line width=.3mm](C2.east) |- (C3.west);
\draw [-latex] [line width=.3mm](C3.east) |- (10.7,-2);
\draw [-latex] [line width=.3mm] (8.9,-3.2) node[right] {$\bm{Y}$} -| (C3.south)  ;
\end{tikzpicture}
}
\end{center}
\vspace{-15pt}
\caption{The Wyner-Ziv coding using real-number codes.}
\label{fig:realDSC}
\end{figure}

We introduce the use of real-number codes in lossy compression of correlated signals. Specifically,
we use DFT codes \cite{Marshall}, a class of real Bose-Chaudhuri-Hocquenghem (BCH) codes, to preform compression.
Similar to error correction in finite fields, the basic
idea of error correcting codes in the real field is to insert
redundancy to a message vector of $k$ samples to convert it to
a codevector of $n$ samples ($n>k$) \cite{Marshall}.
But unlike that, the insertion of
redundancy in the real field is performed before quantization
and entropy coding. The insertion of \textit{soft redundancy} in
the real-number codes has advantages over \textit{hard redundancy}
in the binary field. By using soft redundancy, one can go beyond
quantization error, and thus reconstruct continuous-valued
signals more accurately.
This makes real-number codes more suitable than binary codes
for lossy distributed source coding.

The proposed system is depicted in Fig. \ref{fig:realDSC}. Although it
consists of the same blocks as existing
practical Wyner-Ziv coding scheme \cite{Girod, Xiong}, the order of these blocks is changed here.
That is, we perform Slepian-Wolf coding before quantization.
This change in the order of the DSC and quantization blocks
brings some advantages as described in the following.

\begin{itemize}
  \item \textbf{Realistic correlation model:}
In the existing framework for lossy DSC, correlation between two sources is
modeled after quantization, i.e., in the binary domain. More precisely, correlation
between quantized sources is usually modeled as a BSC, mostly
with known crossover probability. Admittedly though, due to nonlinearity of quantization operation,
correlation between the quantized signals is not known accurately even if it is known in the continuous domain.
This motivates investigating a method that exploits correlation between continuous-valued sources
to perform DSC.

  \item \textbf{Alleviating quantization error:}
In lossy data compression with side information at the decoder, soft redundancy, added by DFT codes, can be used to correct both quantization errors and (correlation) channel errors. The loss due to quantization error thus can be recovered, at least partly if not wholly. More precisely,
if the two sources are exactly the same over a codevector, quantization error can be corrected completely. That is, perfect reconstruction is achieved over corresponding samples.
The loss due to quantization error is decreased even if correlation is not perfect, i.e., when (correlation) channel errors exist.

  \item \textbf{Low-delay communication:}
If communication is subject to low-delay constraints, we cannot use turbo or LDPC codes, as their performance is not satisfactory for short code length.
Whether low-delay requirement exists or not depends on the specific applications. However, even in the applications that low-delay transmission is not imperative, it is sometimes useful to consider low-dimensional systems for their low computational complexity.
\end{itemize}

\section{Encoding and Decoding with BCH-DFT Codes}\label{sec:DFT}

Real BCH-DFT codes, a subset of complex BCH codes \cite{Marshall}, are linear block codes over the real field.
Any BCH-DFT code satisfies two properties.
First, as a DFT code, its parity-check matrix is defined based on the DFT matrix.
Second, similar to other BCH codes, the spectrum of any codevector is zero in a block of $d-1$ cyclically adjacent
components, where $d$ is the designed distance of that code \cite{Blahut2003algebraic}. A real BCH-DFT codes, in addition,
has a generator matrix with real entries, as described below.

\subsection{Encoding}\label{sec:Enc}

An $(n, k)$ real BCH-DFT code is defined by its generator and parity-check matrices.
The generator matrix is given by
\begin{align}
\bm{G}= \sqrt{\frac{n}{k}} \bm{W}_n^H \bm{\Sigma} \bm{W}_k,
\label{eq:G1}
\end{align}

\noindent in which $\bm{W}_k$ and $\bm{W}_n^H$ respectively are the DFT and
IDFT matrices of size $k$ and $n$, and
\mbox{\boldmath$\Sigma$} is an $n \times k$ matrix with $n-k$ zero rows \cite{rath2004subspace,Takos,gabay2007joint,Vaezi2011LS}.  
Particularly, for odd $k$, \mbox{\boldmath$\Sigma$} has exactly $k$ nonzero elements given as $\Sigma_{00}=1$,
$\Sigma_{i,i}=\Sigma_{n-i,k-i}=1$, $i=1:\frac{k-1}{2}$
 \cite{rath2004subspace}, \cite{Takos}. This guarantees the spectrum of any codeword to have $n-k$ consecutive zeros, which is
required for any BCH code \cite{Blahut2003algebraic}.
The parity-check matrix $\bm{H}$, on the other hand, is constructed by using the $n-k$ columns of $\bm{W}_n^H$
corresponding to the $n-k$ zero rows of $\bm{\Sigma}$.
Therefore, due to unitary property of $\bm{W}_n^H$,  $\bm{HG}=\bm{0}$.

In the rest of this paper, we use the term DFT code in lieu of real BCH-DFT code.
Besides, we only consider odd numbers for $k$ and $n$; thus, the error correction capability of the code is
$t=\lfloor \frac{n-k}{2}\rfloor=\frac{n-k}{2}$.
\subsection{Decoding}\label{sec:Dec}
For decoding, we use the extension of the well-known Peterson-Gorenstein-Zierler
(PGZ) algorithm to the real field \cite{Blahut2003algebraic}.
This algorithm, aimed at detecting, localizing, and estimating errors,
works based on the syndrome of error.
We summarize the main steps of this algorithm, adapted for a DFT code of length $n$, in the following.
\begin{enumerate}
  \item Compute vector of syndrome samples
  \item Determine the number of errors $\nu$ by constructing a syndrome matrix and finding its rank
  \item Find coefficients $\Lambda_1, \hdots, \Lambda_\nu$ of error-locating polynomial $\Lambda(x) = \prod_{\substack{i=1}}^{\nu} (1-xX_i)$ whose roots are the inverse of error locations
  \item Find the zeros $X_1^{-1}, \hdots, X_\nu^{-1}$ of $\Lambda(x)$; the errors are then
in locations $i_1, \hdots, i_\nu$ where $X_1 = \alpha^{i_1}, \hdots, X_\nu = \alpha^{i_\nu}$ and  $\alpha= e^{-j\frac{2\pi}{n}}$
  \item Finally, determine error magnitudes by solving a set of linear equations whose constants coefficients are powers of $X_i$.
\end{enumerate}

As mentioned, the PGZ algorithm works based on the syndrome of  error,
which is the syndrome of the received codevector, neglecting quantization.
Let $\bm{r} = \bm{y} + \bm{e}$ be the received
vector, then
\begin{align}
\mbox{\boldmath$ \bm{s}= \bm{Hr}= \bm{H}(\bm{y} + \bm{e})= \bm{He}$},
\label{eq:synd}
\end{align}
where $\bm{s}=[s_1,\, s_2, \hdots, s_{2t}]^T$  is a complex vector of length $n-k$.
In practice however, the received vector is distorted by quantization
($\bm{r} = \bm{\hat y} + \bm{e}, \; \bm{\hat y}=\bm{y}+\bm{q}$) and
 its syndrome is no longer equal to the syndrome of error because
\begin{align}
\bm{\tilde{s}}= \bm{Hr}= \bm{H}(\bm{ y} + \bm{q}+ \bm{e})= \bm{s}_q + \bm{s}_e,
\label{eq:syndq}
\end{align}
where $\bm{s}_q \equiv\bm{Hq}$ and $\bm{s}_e \equiv \bm{He}$.
While the ``exact'' value of errors is determined neglecting quantization,
the decoding becomes an {\it estimation} problem in the presence
of quantization.
Then, it is imperative to modify the PGZ algorithm to detect errors reliably \cite{Blahut2003algebraic,rath2004subspace,Takos,gabay2007joint}.
Error detection, localization, and also estimation
can be largely improved using least squares methods \cite{Vaezi2011LS}.

\subsection{Performance Compared to Binary Codes}\label{sec:Per}

DFT codes by construction are capable of decreasing quantization error.
When there is no error, an $(n, k)$  DFT code brings down the mean-squared
error (MSE), below the level of quantization error, with a factor of $R_c=k/n$ \cite{rath2004frame,goyal2001quantized}.
This is also shown to be valid for channel errors, as long as channel can be modeled as by additive noise.
To appreciate this, one can consider the generator matrix of a DFT code as a tight frame \cite{rath2004frame}; it is
known that frames are resilient to any additive noise, and
tight frames reduce the MSE $k/n$ times \cite{kovacevic2008introduction}.
Hence, DFT codes can result in
a MSE even better than quantization error level whereas the best
possible MSE in a binary code is obviously lower-bounded by quantization error level.


\section{Wyner-Ziv Coding Using DFT Codes} \label{sec:WZ}

The concept of lossy DSC and Wyner-Ziv coding in the real field was
described in Section \ref{sec:sys}. In this section, we use
DFT codes, as a specific means, to do Wyner-Ziv coding in the real field.
This is accomplished by using DFT codes for binning, and
transmitting compressed signal, in the form of either syndrome or parity samples.

Let $\mybf{x}$ be a sequence of i.i.d random variables $x_1x_2 \hdots x_n$,
and $\mybf{y}$ be a noisy version of $\mybf{x}$
such that $y_i=x_i + e_i$, where $e_i$ is continuous, i.i.d., and independent of $x_i$.
Since $\bm{e}$ is continuous, this model precisely captures any variation of $\bm{x}$, so it can
model correlation between $\mybf{x}$ and $\mybf{y}$ accurately.
For example, the Gaussian, Gaussian Bernoulli-Gaussian, and Gaussian-Erasure correlation
channels can be modeled using this model \cite{bassi2008source}.
These correlation models are practically important in video coders
that exploit Wyner-Ziv concepts, e.g., when the decoder builds side information
via extrapolation of previously decoded frames or interpolation of key frames \cite{bassi2008source}.
In this paper, the virtual correlation channel is assumed to be a Bernoulli-Gaussian channel,
inserting at most $t$ random errors in each codeword; thus, $\bm{e}$ is a sparse vector.

\subsection{Syndrome Approach} \label{synd}
\subsubsection{Encoding}
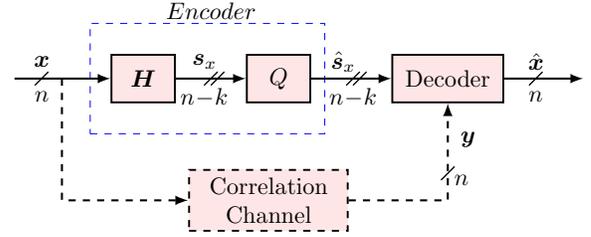
\begin{figure}[!tb]
\begin{center}
\scalebox {.9}{
\begin{tikzpicture}
[auto, block/.style ={rectangle, draw=blue, thick, fill=blue!10, text width=5em,align=center, rounded corners, minimum height=2em},
block2/.style ={rectangle, draw=white, thick, fill=white, text width=2em,align=center, rounded corners, minimum height=1em},
 block1/.style ={rectangle, draw=black, thick, fill=red!10, text width=6.5em,align=center, minimum height=2em},
 line/.style ={draw, thick, -latex',shorten >=2pt},
 cloud/.style ={draw=red, thick, ellipse,fill=red!20,
 minimum height=1em}]
\draw(4,-1)node[block2,text width=6em] {$Encoder$};
\draw(1.5,-1.75)node[block2,text width=1em]{$\bm{x}$};
\draw(1.5,-2.27)node[block2,text width=1em]{$n$};
\draw(3.9,-1.75)node[block2,text width=1em]{$\bm{s}_x$};
\draw(3.9,-2.27)node[block2,text width=2em]{$n-k$};
\draw(5.95,-1.75)node[block2,text width=1em]{$\hat{\bm{s}}_x$};
\draw(6.1,-2.27)node[block2,text width=2em]{$n-k$};
\draw(8.8,-1.75)node[block2,text width=1em]{$\hat{\bm{x}}$};
\draw(8.8,-2.27)node[block2,text width=1em]{$n$};
\draw(7.8,-2.9)node[block2,text width=1em]{$\bm{y}$};
\draw(7.7,-3.5)node[block2,text width=1em]{$n$};
\draw [line width=0.2mm](1.4,-2.1) -- (1.6,-1.9);
\draw [line width=0.2mm](3.9,-2.1) -- (4.1,-1.9);
\draw [line width=0.2mm](4,-2.1) -- (4.2,-1.9);
\draw [line width=0.2mm](6,-2.1) -- (6.2,-1.9);
\draw [line width=0.2mm](6.1,-2.1) -- (6.3,-1.9);
\draw [line width=0.2mm](8.7,-2.1) -- (8.9,-1.9);
\draw [line width=0.2mm](7.4,-4.1+.6) -- (7.6,-3.9+.6);
\draw (3,-2) node[block1, text width=2em] (C) {$\bm{H}$};
\draw (5,-2) node[block1, text width=2em] (C1) {$Q$};
\draw (7.5,-2) node[block1 , text width=4em] (C2) {$\mathop{\mathrm{Decoder }}$};
\draw (4.85,-3.8) node[block1,dashed, text width=6em] (C3) {$\mathop{\mathrm{Correlation }}$ \\ $\mathop{\mathrm{ Channel}}$};
\draw[blue,dashed] ($(C.north west)+(-0.3,0.45)$) rectangle ($(C.south east)+(2.2,-0.45)$)  ;
\draw [-latex] [line width=.3mm]  (1.1,-2) |- (C.west) ;
\draw [-latex] [line width=.3mm] (C.east) |- (C1.west) ;
\draw [-latex] [line width=.3mm](C1.east) |- (C2.west);
\draw [-latex] [line width=.3mm](C2.east) |- (9.5,-2) ;
\draw [-latex] [line width=.3mm,dashed](1.8,-2) |- (C3.west);
\draw [-latex] [line width=.3mm,dashed](C3.east) -| (C2.south);
\end{tikzpicture}
}
\end{center}
\caption{The Wyner-Ziv coding using DFT codes: Syndrome approach.}
\label{fig:WZsynd}
\vspace{-.3cm}
\end{figure}

Given \mybf{H}, to compress an arbitrary sequence of data samples, we
multiply it with \mybf{H} to find the corresponding
syndrome samples \bfsub{s}{x}\mybf{=Hx}. The syndrome
is then quantized (\bfsubp{\hat s}{x}{}$\;=\;$\bfsubp{s}{x}{}$\;+\; \mybf{q}$), and transmitted
over a noiseless digital communication system, as shown in
Fig. \ref{fig:WZsynd}. Note that
 \mbox{\boldmath$s$}$_x$, \bfsub{\hat s}{x} are both complex vectors of length $n-k$.

\subsubsection{Decoding}

The decoder estimates the input sequence from the received syndrome and side
information $\bm{y}$. To this end,
it needs to evaluate the syndrome of channel (correlation) errors. This
can be simply done by subtracting the received syndrome from syndrome of
side information. Then, neglecting quantization, we obtain,
\begin{align}
\mbox{\bfsubp{s}{e}{}$ \; = \; $\bfsubp{s}{y}{}$\;- \;$\bfsubp{s}{x}{}},
\label{eq:synd5}
\end{align}
and \bfsubp{s}{e}{} can be used to precisely estimate the error vector,
as described in Section \ref{sec:Dec}.
In practice, however, the decoder knows \bfsubp{\hat s}{x}{}$\;=\;$\bfsubp{s}{x}{}$\;+\; \mybf{q}$
rather than \mbox{\boldmath$ s_x$}. Therefore, only a distorted syndrome
of error is available, i.e.,
\begin{align}
\mbox{\bfsubp{\tilde{s}}{e}{}$ \; = \; $\bfsubp{s}{y}{}$ \; - \;$\bfsubp{\hat s}{x}{}$ \; = \; $\bfsubp{s}{e}{}$ \; - \; $ \mybf{q}}.
\label{eq:synd6}
\end{align}
Hence, using the PGZ algorithm, error correction is accomplished based on \eqref{eq:synd6}.
Note that, having computed the syndrome of error, decoding algorithm
in DSC using DFT codes is exactly the same as that in the channel coding problem.
This is different from DSC techniques in the binary field which
usually require a slight modification in the corresponding
channel coding algorithm to customize for DSC.

\subsection{Parity Approach} \label{par}

Syndrome-based Wyner-Ziv coding is straightforward but not very efficient
because, in a real DFT code, syndrome samples are complex numbers.
This means that to transmit each sample we need to send two real numbers,
one for the real part and one for the imaginary part. Thus, the compression ratio,
using an $(n, k)$ DFT code, is $\frac{n}{2(n-k)}$ whereas it is $\frac{n}{n-k}$
for a similar binary code. This also imposes a constraint
on the rate of code, i.e., $n<2k$ or $R_c>\frac{1}{2}$, since
otherwise there is no compression.
In the sequel, we explore parity-based approach to the Wyner-Ziv coding.

\subsubsection{Encoding}

\begin{figure}[!tb]
\begin{center}
\scalebox {.9}{
\begin{tikzpicture}
[auto, block/.style ={rectangle, draw=blue, thick, fill=blue!10, text width=5em,align=center, rounded corners, minimum height=2em},
 block1/.style ={rectangle, draw=black, thick, fill=red!10, text width=6.5em,align=center, minimum height=2em},
 block2/.style ={rectangle, draw=white, thick, fill=white, text width=2em,align=center, rounded corners, minimum height=1em},
 line/.style ={draw, thick, -latex',shorten >=2pt},
 line1/.style ={draw, thick, -.latex',shorten >=2pt},
 cloud/.style ={draw=red, thick, ellipse,fill=red!20,
 minimum height=1em}]
\draw(4,-1)node[block2,text width=6em] {$Encoder$};
\draw(1.5,-1.75)node[block2,text width=1em]{$\bm{x}$};
\draw(1.5,-2.27)node[block2,text width=1em]{$k$};
\draw(3.9,-1.75)node[block2,text width=1em]{$\bm{p}$};
\draw(3.9,-2.27)node[block2,text width=2em]{$n-k$};
\draw(5.95,-1.75)node[block2,text width=1em]{$\hat{\bm{p}}$};
\draw(6.1,-2.27)node[block2,text width=2em]{$n-k$};
\draw(8.8,-1.75)node[block2,text width=1em]{$\hat{\bm{x}}$};
\draw(8.8,-2.27)node[block2,text width=1em]{$k$};
\draw(7.8,-2.9)node[block2,text width=1em]{$\bm{y}$};
\draw(7.7,-3.5)node[block2,text width=1em]{$k$};
\draw (3,-2) node[block1, text width=2em] (C) {$\bm{G}_{\mathrm{sys}}$};
\draw (5,-2) node[block1, text width=2em] (C1) {$Q$};
\draw (7.5,-2) node[block1 , text width=4em] (C2) {$\mathop{\mathrm{Decoder}}$};
\draw (4.85,-3.8) node[block1, dashed, text width=6em] (C3) {$\mathop{\mathrm{Correlation }}$ \\ $\mathop{\mathrm{ Channel}}$};
\draw[blue,dashed]  ($(C.north west)+(-0.3,0.45)$) rectangle ($(C.south east)+(2.2,-0.45)$) ;
\draw [-latex] [line width=.3mm]  (1.1,-2)  |- (C.west) ;
\draw [-latex] [line width=.3mm] (C.east) |- (C1.west) ;
\draw [-latex] [line width=.3mm](C1.east) |- (C2.west);
\draw [-latex] [line width=.3mm](C2.east) |- (9.5,-2);
\draw [-latex] [dashed, line width=.3mm](1.8,-2) |- (C3.west);
\draw [-latex] [dashed, line width=.3mm](C3.east) -| (C2.south);
\draw [line width=0.2mm](1.4,-2.1) -- (1.6,-1.9);
\draw [line width=0.2mm](3.9,-2.1) -- (4.1,-1.9);
\draw [line width=0.2mm](4,-2.1) -- (4.2,-1.9);
\draw [line width=0.2mm](6,-2.1) -- (6.2,-1.9);
\draw [line width=0.2mm](6.1,-2.1) -- (6.3,-1.9);
\draw [line width=0.2mm](8.7,-2.1) -- (8.9,-1.9);
\draw [line width=0.2mm](7.4,-4.1+.6) -- (7.6,-3.9+.6);
\end{tikzpicture}
}
\end{center}
\caption{The Wyner-Ziv coding using DFT codes: Parity approach.}
\label{fig:WZparity}
\end{figure}
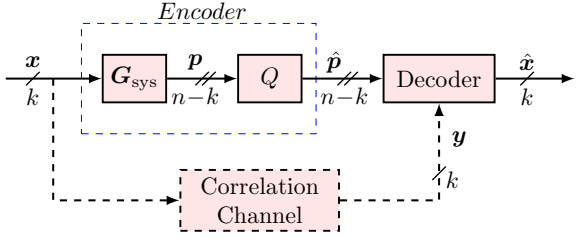

To compress $\bm{x}$, the encoder generates the corresponding
parity sequence $\bm{p}$ with $n-k$
samples. The parity is
then quantized and transmitted, as shown in
Fig.~\ref{fig:WZparity}, instead of transmitting the input data.
The first step in parity-based system is to find the systematic
generator matrix, as $\bm{G}$ in \eqref{eq:G1}
is not in the systematic form.
Let $\bm{H}$ be partitioned as $\bm{H} = [\bm{H}_1 \;|\; \bm{H}_2]$, where $\bm{H}_1 $
is a matrix of size $n-k\times k$, and $\bm{H}_2 $
is a square matrix of size $n-k$.
Since $\bm{H}_2$ is  a Vandermonde matrix, $\bm{H}_2^{-1}$
exist and we can write
\begin{align}
\bm{H}_{\mathrm{sys}}= \bm{H}_2^{-1} \bm{H}= [\bm{P} \;|\; \bm{I}_{2t}],
\label{eq:synd5}
\end{align}
in which $\bm{P}= \bm{H}_2^{-1}\bm{H}_{1}$ is an $(n-k)\times k$ matrix, and
$\bm{I}_{2t}$ is an identity matrix of size $2t$.

The systematic generator matrix corresponding to $ \bm{H}_{\mathrm{sys}}$
is given by
\begin{align}
\bm{G}_{\mathrm{sys}} &= \left[\begin{array}{c}
      \bm{I}_k \\ 
      -\bm{P}
    \end{array}\right]
    =\left[\begin{array}{c}
      \bm{I}_k \\ 
      -\bm{H}_2^{-1} \bm{H}_1
    \end{array}\right].
\label{eq:Gsys}
\end{align}
Clearly, $\bm{H}_{\mathrm{sys}}\bm{G}_{\mathrm{sys}}=\bm{0}$. It is also easy to check that
\begin{align}
 \bm{H}\bm{G}_{\mathrm{sys}}=\bm{0}.
\label{eq:null}
\end{align}
Therefore, we do not need to calculate $\bm{H}_{\mathrm{sys}}$ and the same
parity-check matrix $\bm{H}$ can be used for decoding in the
parity approach.

An even easier way to come up with systematic generator matrix is
to partition $\bm{G}$ as $\left[\begin{array}{c}
      \bm{G}_{1} \\ 
      \bm{G}_{2}
    \end{array}\right]$  where $\bm{G}_{1}$ is a square matrix of size $k$.
Then, from $\bm{H}\bm{G}=\bm{0}$ and the fact that $\bm{H}_{2}$ is invertible one can see $\bm{G}_{2}=-\bm{H}_2^{-1} \bm{H}_1\bm{G}_{1}$; thus, we have
 \begin{align}
  \bm{G} &= \left[\begin{array}{c}
      \bm{G}_{1} \\ 
      \bm{G}_{2}
    \end{array}\right]
    =\left[\begin{array}{c}
      \bm{I}_k \\ 
      -\bm{H}_2^{-1} \bm{H}_1
    \end{array}\right]\bm{G}_{1}.
\label{eq:Gsys2}
\end{align}
Note that $\bm{G}_{1}$ is invertible because using \eqref{eq:G1} any $k\times k$ submatrix of
$\bm{G}$ can be represented as product of a Vandermonde matrix and the DFT matrix $\bm{W}_{k}$.
This is also proven using a different approach in \cite{rath2004frame}, where it is shown that any subframe of $\bm{G}$
is a frame and its rank is equal to $k$.
Hence, since $\bm{G}_{1}$ is invertible, the systematic generator matrix is given by
  \begin{align}
  \bm{G}_{\mathrm{sys}} =\bm{G}\bm{G}_{1}^{-1}.
\label{eq:Gsys2}
\end{align}
Again $ \bm{H}\bm{G}_{\mathrm{sys}}=\bm{0}$
because $\bm{H}\bm{G}=\bm{0}$. Therefore, the same
parity-check matrix $\bm{H}$ can be used for decoding in the
parity approach.
It is also easy to see that $\bm{G}_{\mathrm{sys}}$ is a real matrix.
The question that
remains to be answered is whether $\bm{G}_{\mathrm{sys}}$ corresponds to a BCH code?
To generate a BCH code, $\bm{G}_{\mathrm{sys}}$ must have $n-k$ consecutive
zeros in the transform domain. $\bm{W}_n\bm{G}_{\mathrm{sys}}=(\bm{W}_n\bm{G})\bm{G}_{1}^{-1}$, the Fourier transform of this matrix
 satisfies this condition because $\bm{W}_n\bm{G}$,
the Fourier transform of original matrix, satisfies that.

Note that, since parity samples, unlike syndrome samples, are real numbers,
using an $(n, k)$ DFT code a compression ratio  of $\frac{k}{n-k}$
is achieved. Obviously, a compression ratio  of $\frac{n}{n-k}$ is
achievable if we use a $(2n-k, n)$ DFT code.

\subsubsection{Decoding}

A parity decoder estimates the input sequence from the received parity and side information $\bm{y}$.
Similar to the syndrome approach, at the decoder, we need to find the syndrome
of channel (correlation) errors.
To do this, we append the parity to the side information and
form a vector of length $n$ whose syndrome, neglecting quantization, is equal to the
syndrome of error. That is,
 \begin{align}
\bm{z}=\left[ \begin{array}{cc} \bm{y}   \\
 \bm{p} \\  \end{array} \right]
 =\left[ \begin{array}{cc} \bm{x}   \\
 \bm{p}  \end{array}\right]  + \left[ \begin{array}{cc} \bm{e}_k   \\
 \bm{0}\end{array}\right]= \bm{G}_{\mathrm{sys}}\bm{x}+\bm{e}_n,
 \end{align}
hence,
\begin{align}
 \bm{s}_z=\bm{s}_e.
 \end{align}
Similarly, when quantization is involved ($\bm{\hat p}=\bm{p}+\bm{q}$), we get
  \begin{align}
 \bm{\tilde{z}} =\left[ \begin{array}{cc} \bm{y}   \\
 \bm{\hat p} \\  \end{array} \right]
 =\bm{z} +  \left[ \begin{array}{cc} \bm{0}   \\
\bm{q}\end{array}\right]= \bm{G}_{\mathrm{sys}}\bm{x}+\bm{e}_n+\bm{q}_n,
 \end{align}
and
\begin{align}
 \bm{s}_{\tilde{z}}=\bm{s}_e+\bm{s}_q,
 \end{align}
 in which, $\bm{s}_q \equiv \bm{Hq}_n$.
Therefore, we obtain a distorted version of error syndrome.
In both cases, the rest of the algorithm, which is based on the
syndrome of error, is similar to that in the channel coding problem using DFT codes.

\subsection{Comparison Between the Two Approaches} \label{comp}
As we saw earlier, using an $(n, k)$ code
the compression ratio in the syndrome and parity approaches, respectively, is $\frac{n}{2(n-k)}$
and $\frac{k}{n-k}$. Hence, the parity approach is $2k/n=2R_c >1$ times
more efficient than the syndrome approach. Conversely, we can find two different codes that result in
same compression ratio, say $\frac{n}{n-k}$. We know that in the parity approach,
a $(2n-k, n)$ code can be used for this matter. It is also easy to
verify that, in the syndrome approach, a code with rate $R_c = \frac{n+k}{2n}$ results
in the same compression. For odd $n$ and $k$, the $(n, \frac{n+k}{2})$
DFT code gives the desired compression ratio.
Thus, for a given compression ratio the parity approach implies a code with smaller rate compared to the code
required in the syndrome approach.

\section{Simulation Results}
\label{sec:sum}

\begin{figure}
  \centering
 \includegraphics [scale=0.5] {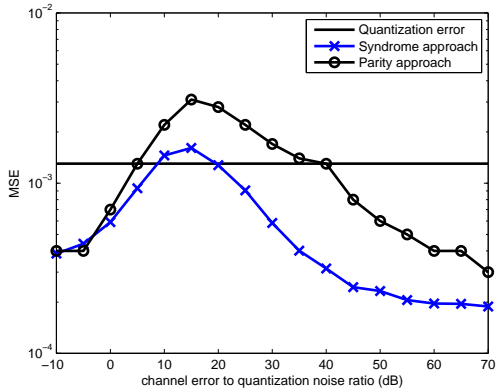}
  \caption{ Reconstruction error in the syndrome and parity approaches, using a $(7, 5)$ DFT code in Fig. 2, 3.
  For both schemes, the virtual correlation channel inserts one error
  at each channel error to quantization noise ratio. }
  \label{fig:syndQ}
    \vspace{-.05cm}
\end{figure}

We evaluate the performance of the proposed systems using
a Gauss-Markov source with zero mean, unit variance, and
correlation coefficient 0.9; the effective range of the input
sequences is thus $[-4, 4]$.
The sources sequences are binned using a $(7, 5)$ DFT code.
The compressed vector, either syndrome or parity, is then
quantized with a 6-bit uniform quantizer, and
transmitted over a noiseless communication media.
The correlation channel randomly inserts one error $(t=1)$,
generated by a Gaussian distribution.
The decoder localizes and decodes errors.
We compare the MSE between transmitted and reconstructed
codevectors, to measurers end to end
distortion. In all simulations, we use 20,000
input frames for each channel-error-to-quantization-noise ratio (CEQNR).
We vary the CEQNR and plot the resulting MSE.
The result are presented in Fig.~\ref{fig:syndQ}, and compared
against the quantization error level in the existing lossy DSC methods.

\begin{figure}
  \centering
  \includegraphics [scale=0.5] {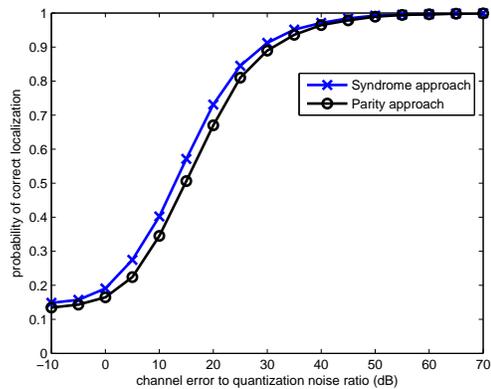}
  \caption{ Relative frequency of correct localization of correlation channel error in the syndrome and parity approaches, using a $(7, 5)$ DFT code.
   }
  \label{fig:PoE}
\end{figure}

It can be observed that the MSE in the syndrome approach is lower than
quantization error except for a small range of CEQNR. Similarly,
in the parity approach, the MSE is less than
quantization error for a wide range of CEQNR. Note that in lossy DSC using binary
codes, the MSE can be equal to quantization error only if the probability
of error is zero. The performance of both algorithms improves as CEQNR is very high.
This improvement is due to
better error localization, since the higher the CEQNR the better the error
localization, as shown in Fig.~\ref{fig:PoE} and \cite{rath2004subspace}.
At very low CEQNRs, although error localization is poor, the MSE is still
very low because, compared to quantization error,
the errors are so small that the algorithm may
localize and correct some of quantization errors instead.
Additionally, reconstruction error is always reduced
with a factor of $R_c=k/n$, in an $(n,k)$ DFT code.

In terms of compression, the parity approach is $2R_c=\frac{10}{7}$ times more
efficient than the syndrome approach, as discussed earlier in Section~\ref{comp}.
Not surprisingly though, the performance of the parity approach is
not as good as that of the syndrome approach, because
it contains fewer redundant samples. On top of that, in this simulation,
$1/5$ of samples are corrupted in the parity approach
while this figure is $1/7$ for the syndrome approach.
The parity approach, however, suffers from the fact that
dynamic range of parity samples, generated by \eqref{eq:Gsys2},
could be much higher than that of syndrome
samples as $t$ increases. This implies more precision bits to achieve the same accuracy.
Finally, it is worth mentioning that
when data and side information are the same over a block of code,
reconstruction error becomes zero in both approaches.

\section{Conclusions}\label{sec:con}
We have introduced a new framework for distributed lossy source coding in general,
and Wyner-Ziv coding specifically. The idea
is to do binning before quantizing the continuous-valued signal, as opposed to the conventional
approach where binning is done after quantization.
By doing binning in the real field, the virtual correlation channel
can be modeled more accurately, and
quantization error can be corrected when there is no error.
In the new paradigm,
Wyner-Ziv coding is realized by cascading a Slepian-Wolf encoder
with a quantizer. We employ real BCH-DFT codes to do the Slepian-Wolf in the real field.
At the decoder, by introducing both syndrome-based and parity-based systems,
we adapt the PGZ decoding algorithm accordingly.
From simulation results, we conclude that our
systems, specifically with short codes, can improve the reconstruction error, so that they may become viable in real-world scenarios, where low-delay communication
is required.

\end{document}